\documentclass[aps,prl,showpacs,superscriptaddress,floatfix]{revtex4}
\usepackage{epsfig}\usepackage{amsmath}
\usepackage{amsmath,amsfonts,amssymb,graphics,graphicx,epsfig,color,times,bbm}
\newcommand{\id}{{\sf 1 \hspace{-0.3ex} \rule{0.1ex}{1.52ex}\rule[-.01ex]{0.3ex}{0.1ex}}}

\begin{document}

\preprint{APS/123-QED}

\title{Effective cross-Kerr nonlinearity and robust phase gates with trapped ions}

\author{F. L. Semi\~ao}
\email{semiao@ifi.unicamp.br}
\author{A. Vidiella-Barranco}%
\email{vidiella@ifi.unicamp.br}
\affiliation{%
Instituto de F\'\i sica ``Gleb Wataghin'' - Universidade Estadual de
Campinas, 13083-970 Campinas, S\~ao Paulo, Brazil}
\date{\today}

\begin{abstract}
We derive an effective Hamiltonian that describes a cross-Kerr type
interaction in a system involving a two-level trapped ion coupled to
the quantized field inside a cavity. We assume a large detuning
between the ion and field (dispersive limit) and this results in an
interaction Hamiltonian involving the product of the (bosonic) ionic
vibrational motion and field number operators. We also demonstrate
the feasibility of operation of a phase gate based on our
hamiltonian. The gate is insensitive to spontaneous emission,
an important feature for the practical implementation of
quantum computing.
\end{abstract}

\pacs{03.67.Lx, 42.65.Hw, 32.80.Qk}
\maketitle

Nonlinear effects are specially relevant to quantum optics as well
as to applications in quantum information and computation. In
particular, the cross-Kerr coupling is at the basis of various
schemes, such as quantum non-demolition measurements
\cite{imoto1,poizat1}, a Fock state synthesizer \cite{dariano1},
entanglement purification \cite{zoller1}, teleportation
\cite{tombesi1}, quantum codewords \cite{tombesi2}, a proposal of
realization of a CNOT gate \cite{munro1}, and Bell-state detection
\cite{barret}, for instance. Large cross-Kerr nonlinearities are
usually required for the actual implementation of those schemes, but
the values of the Kerr coefficients in the available nonlinear media
are in general relatively small, although some progress based on
atomic gases has been achieved \cite{imamoglu,hau,tombesi3}. It is
therefore of interest to seek alternative ways of obtaining
nonlinear interactions of that type. The cross-Kerr interaction may
involve the effective coupling of two modes of the field
\cite{tombesi3}, or even two sub-systems of different nature, such
as an atom and a field \cite{brauns1}.

Thinking about specific applications, and given the extraordinary
development of the field of quantum computing, it is
important to identify physical systems and situations that could be
suitable for the realization of efficient quantum logic gates. A
rather interesting physical system comprises a single trapped ion
coupled to the quantized field \cite{first}. That configuration
forms a tripartite quantum system: the internal electronic degrees
of freedom; the quantized center-of-mass motion; and the quantum
cavity field. Several aspects of that system connected to quantum
information and computation have been already investigated, e.g., a
scheme for generation of entangled states involving the ionic
vibrational motion and the quantized field \cite{ours1}, as well as
an alternative proposition of implementation of a CNOT gate
\cite{ours2}. As a matter of fact, the possibility of coherent
manipulation of trapped ions (qubits suitable for storing quantum
information) coupled to photons (qubits suitable for carrying
quantum information) \cite{blatt1} opens new possibilities for
quantum information processing \cite{blatt2}.

In this paper we derive, from a general hamiltonian involving the
oscillating ion interacting with the quantized field, an effective
Hamiltonian which couples the quantized field and the ionic motion
in a way similar to what it is found in nonlinear optics. More
specifically, we obtain a coupling which is basically the
product of bosonic number operators referring to the vibrational
motion ($\hat{a}^{\dagger}\hat{a}$) and the quantized field
($\hat{b}^{\dagger}\hat{b}$), yielding an interaction hamiltonian of
the form $\hat{H}_{int}=\lambda\hat{a}^{\dagger}\hat{a}
\hat{b}^{\dagger}\hat{b}$. In the second part of our paper, we
suggest an application of our Hamiltonian for quantum computing,
presenting a proposal of implementation of a quantum phase gate. In
our scheme, the ion is initially prepared in its ground
(electronic) state $|g\rangle$, remaining in that state during the
gate operation, which means that the gate will be quite robust
against atomic spontaneous emission.

We consider a two level ion (with atomic frequency
$\omega_{\rm{a}}$) trapped in a harmonic potential (frequency $\nu$)
inside a high finesse cavity, and coupled to a single mode
(frequency $\omega_{\rm{c}}$) of the quantized cavity field. The ion
has its motion well cooled down and therefore it should described
quantum mechanically. The Hamiltonian for that system reads
\cite{first}
\begin{equation}
    H=\nu \hat{a}^\dag \hat{a}+\omega_{\rm{c}} \hat{b}^\dag \hat{b}+
    \frac{1}{2}\omega_{\rm{a}}\hat{\sigma}_{\rm{z}}+g(\hat{\sigma}_+\hat{b}+\hat{b}^\dag
    \hat{\sigma}_-)\cos\eta(\hat{a}^\dag+\hat{a}),\label{h}
\end{equation}
where $\hat{a}^{\dagger}(\hat{a})$ are the creation(annihilation)
operators relative to the excitations of the center-of-mass
oscillatory motion, $\hat{b}^{\dagger}(\hat{b})$ are the creation
(annihilation) operators of photons in the field mode, and $g$ is
the ion-field coupling constant. The operators
$\hat{\sigma}_{\rm{z}}$, $\sigma_+$ ($\hat{\sigma}_-$) are the
population difference and excitation (de-excitation) atomic
operators, respectively. In order to derive a Kerr-like interaction
we need to perform a couple of approximations. By one side, we note
that the factor $\cos\eta(\hat{a}^\dag+\hat{a})$ in Hamiltonian
($\ref{h}$) may be expanded in powers of the operators $\hat{a}$ and
$\hat{a}^\dag$ giving then rise to different processes changing
either energy or phase of the center-of-mass motion. It is then
mandatory to get rid of the energy changing terms, keeping just
terms proportional to powers of the number operator
$\hat{a}^\dag\hat{a}$. In the interaction picture, the Hamiltonian
($\ref{h}$) is given by
\begin{equation}
\hat{H}_I=g(\hat{\sigma}_+\hat{b}e^{i\Delta
t}+\hat{\sigma}_-\hat{b}^\dag e^{-i\Delta t})\cos\eta(\hat{a}^\dag
e^{i\nu t} +\hat{a}e^{-i\nu t}),
\end{equation}
where we defined $\Delta=\omega_{\rm{a}}-\omega_{\rm{c}}$. Making
use of the relation
$e^{\hat{A}+\hat{B}}=e^{-[\hat{A},\hat{B}]/2}e^{\hat{A}}e^{\hat{B}}$,
valid for $[\hat{A},\hat{B}]=\gamma$, where $\gamma$ is an arbitrary
complex number, we have
\begin{equation}
\hat{H}_{I}=g\hat{\sigma}_+\hat{b}\sum_{\alpha,\beta}F(\hat{a}^\dag,\hat{a};\alpha,\beta)e^{i[\Delta+\nu(\alpha-\beta)]
t}+g\hat{\sigma}_-\hat{b}^\dag\sum_{\alpha,\beta}F(\hat{a}^\dag,\hat{a};\alpha,\beta)e^{-i[\Delta+\nu(\alpha-\beta)]
t},\label{intint}
\end{equation}
where we defined the function
\begin{equation}
F(\hat{a}^\dag,\hat{a};\alpha,\beta)=\frac{e^{-\eta^2/2}}
{2\,\alpha!\beta!}[(i\eta)^{\alpha+\beta}+(-i\eta)^{\alpha+\beta}]a^\dag{}^\alpha
a^\beta.
\end{equation}
Now, we perform the first approximation. By analyzing the temporal
dependence in Hamiltonian (\ref{intint}), we may carefully choose
the frequencies of the system in order to perform a suitable
rotating wave approximation. In this approximation, rapidly
oscillating terms are dropped out and just the slow ones are kept.
The terms in (\ref{intint}) oscillate in time either as $e^{\pm
i\Delta t}$ or $e^{i\pm(\Delta+k\nu)t}$, with $k$ integer. In the
regime $\Delta\ll \nu$ and $\Delta\neq k\nu$, the terms oscillating
with lower frequencies are those proportional to $e^{\pm i\Delta
t}$. This occurs if $\alpha=\beta$, or
\begin{equation}
F(\hat{a}^\dag,\hat{a};\alpha,\alpha)=e^{-\eta^2/2}\frac{(-1)^\alpha\eta^{2\alpha}a^\dag{}^\alpha
a^\alpha.} {\alpha!^2},
\end{equation}
and then
\begin{equation}
\hat{H}=\nu\hat{a}^\dag\hat{a}+\omega_c\hat{b}^\dag\hat{b}+\frac{\omega_a}{2}\hat{\sigma}_z+
g(\hat{\sigma}_+\hat{b}+\hat{\sigma}_-\hat{b}^\dag)f(\hat{a}^\dag\hat{a}),
\end{equation}
with
\begin{equation}
f(\hat{a}^\dag\hat{a})=e^{-\eta^2/2}:J_0(\mathfrak{n}):\label{f},
\end{equation}
where $:J_0(\mathfrak{n}):$ is just the normally
ordered zeroth order Bessel function of the first kind
\cite{bana}, with $\mathfrak{n}\equiv
2\eta\sqrt{\hat{a}^\dag\hat{a}}$. We note the presence of $\hat{b}$
and $\hat{b}^\dag$, which must be replaced by terms proportional to
$\hat{b}^\dag\hat{b}$. This may be accomplished by considering the
large detuning limit $\Delta\gg g$. We proceed and write down the
Heisenberg equations of motion for the atomic and field transition
operators. From (\ref{h}) one obtains
\begin{eqnarray}
i\frac{d\hat{\sigma}_+}{dt}&=&-\omega_a\sigma_++g\hat{b}^\dag\hat{\sigma}_zf(\hat{a}^\dag\hat{a})\\
i\frac{d\hat{b}}{dt}&=&\omega_c\hat{b}+g\hat{\sigma}_-f(\hat{a}^\dag\hat{a})\\
i\frac{d\hat{a}}{dt}&=&\nu\hat{a}+g(\hat{\sigma}_+\hat{b}+\hat{\sigma}_-\hat{b}^\dag)
\frac{\partial}{\partial\hat{a}^\dag}f(\hat{a}^\dag\hat{a})\label{a}.
\end{eqnarray}
It is now suitable to move to a new frame using the transformation
\begin{eqnarray}
\hat{\sigma}_+&\rightarrow&\hat{\sigma}_+e^{i\omega_a t}\\
\hat{b}(t)&\rightarrow&\hat{b}e^{-i\omega_c t}\\
\hat{a}(t)&\rightarrow&\hat{a}e^{-i\nu t},
\end{eqnarray}
so that the above equations of motion can be rewritten as
\begin{eqnarray}
i\frac{d\hat{\sigma}_+}{dt}&=&g\hat{b}^\dag\hat{\sigma}_zf(\hat{a}^\dag\hat{a})e^{-i\Delta t}\label{sigmais}\\
i\frac{d\hat{b}}{dt}&=&g\hat{\sigma}_-f(\hat{a}^\dag\hat{a})e^{-i\Delta t}\label{b}\\
i\frac{d\hat{a}}{dt}&=&g(\hat{\sigma}_+\hat{b}e^{i\Delta
t}+\hat{\sigma}_-\hat{b}^\dag e^{-i\Delta t})e^{-i\nu
t}\frac{\partial}{\partial\hat{a}^\dag}f(\hat{a}^\dag\hat{a}).
\end{eqnarray}
In the large detuning limit, i.e. for $\Delta=\omega_a - \omega_c
\gg g$, we may perform the second order approximation which consists
in assuming that the operators vary slowly when compared to the
relevant frequencies involved. This allows us to integrate
(\ref{sigmais}) obtaining
\begin{equation}
\hat{\sigma}_+=\frac{ge^{-i\Delta
t}}{\Delta}\hat{b}^\dag\hat{\sigma}_zf(\hat{a}^\dag\hat{a}).\label{sigeff}
\end{equation}
After substituting (\ref{sigeff}) into (\ref{b}), it results
\begin{equation}
i\frac{d\hat{b}}{dt}=\frac{g^2}{\Delta}f^2(\hat{a}^\dag\hat{a})\hat{b}\hat{\sigma}_z.
\end{equation}
This equation of motion can be obtained from the effective
Hamiltonian
\begin{equation}
\hat{H}_{{\rm{eff}}}=\nu\hat{a}^\dag\hat{a}+\omega_c\hat{b}^\dag\hat{b}+\frac{\omega_a}{2}\hat{\sigma}_z+
\frac{g^2}{\Delta}f^2(\hat{a}^\dag\hat{a})(\hat{\sigma}_+\hat{\sigma}_-\hat{b}\hat{b}^\dag
-\hat{\sigma}_-\hat{\sigma}_+\hat{b}^\dag \hat{b}).\label{eff}
\end{equation}

One could wonder now what is the validity of such effective
Hamiltonian. Its limit of application may be found by using ordinary
time-dependent perturbation theory. Once the effective Hamiltonian
does not allow any transition between bare states we want the
transitions
$|g,n_c,m_{\rm{v}}\rangle\rightarrow|e,(n-1)_c,m^{'}_{\rm{v}}\rangle$
to be unlikely to happen. Mathematically, the transition probability
is given by
\begin{eqnarray}
P(t)=|\langle e,
(n-1)_c,m^{'}_{\rm{v}}|\hat{U}(t)|g,n_c,m_{\rm{v}}\rangle|^2,
\end{eqnarray}
where $\hat{U}(t)$ is the time evolution operator. In first order,
\begin{equation}
\hat{U}(t)=\id-\imath\int_0^t\hat{H}(t^{'})\,dt^{'},
\end{equation}
being $\hat{H}(t)$ the original Hamiltonian (\ref{h}) in the
interaction picture. After a lengthy calculation, the
wanted transition probability is found to be
\begin{eqnarray}
P(t)=\left(\frac{g\,
e^{-\eta^2/2}}{\Delta+\nu(m-m^{'})}\right)^2A(t)+\left(\frac{g\,
e^{-\eta^2/2}}{\Delta-\nu(m-m^{'})}\right)^2B(t),
\end{eqnarray}
where $A(t)$ and $B(t)$ are oscillating functions of time. It is
then clear that the effective Hamiltonian is to be safely applied
when the conditions $\Delta\neq k\nu \ \ (k = 0,\pm 1, \pm
2,\ldots)$ and $g\ll\Delta $ are fulfilled, as well as
$\Delta\ll\nu$ (for $m\neq m^{'}$).

The Hamiltonian in eq. (\ref{eff}) has novel and interesting
features. It involves products of the number field operator $b^\dag
b$ with the vibrational number operator $a^\dag a$, within the
Bessel function $J_0$, characterizing a cross-Kerr type interaction
between two bosons. The Stark shifts resulting from the large
detuning condition we have assumed, gave rise to the interaction
terms characteristic of the cross-Kerr interaction. The key point
here is that the vibrational number operator $a^\dag a$ comes out
multiplying the light shifts. For a sufficiently small value of the
Lamb-Dicke parameter $\eta$ (Lamb-Dicke regime), we may truncate the
Bessel function in order to obtain a term proportional to
$\eta^2a^\dag a$. From (\ref{f}) we note that
\begin{eqnarray}
f^2(\hat{a}^\dag\hat{a})\approx 1-\eta^2-2\eta^2\hat{a}^\dag\hat{a},
\end{eqnarray}
and then, if the ion is initially prepared (internal levels) in its
ground state $|g \rangle$, we may rewrite (\ref{eff}) in that
approximation as
\begin{eqnarray}
\hat{H}_{{\rm{eff}}}^{{\rm{LD}}}=\nu\hat{a}^\dag\hat{a}+\tilde{\omega}_c\hat{b}^\dag\hat{b}+
\lambda\hat{a}^\dag\hat{a}\hat{b}^\dag\hat{b}, \label{gs}
\end{eqnarray}
where $\tilde{\omega}_c=\omega_c+\eta^2g^2/\Delta-g^2/\Delta$ is a
shift in the field frequency and the coupling constant between the
bosonic modes is $\lambda=2\eta^2g^2/\Delta$. The interaction term
of (\ref{gs}) is precisely the sought Kerr-type interaction in its
well known form. By careful choice of the $\eta$, $g$ and $\Delta$
it is possible to obtain a considerable non-linearity. According to
\cite{blatt3}, a realistic choice for the ion-field coupling could
be $g=2\pi\times1.51$MHz. In the Lamb-Dicke regime, we may consider
$\eta=0.05$ and for the high detuning limit $\Delta=10g$ is also
adequate. These values lead to $\lambda\approx 5$ kHz.

Now, we would like to show the phase gate operation using our
effective hamiltonian. We would like to point out that other few
proposals may be found in the literature, involving trapped ions
\cite{solano1,ours2}, cavity QED \cite{scully1} or optical lattices
\cite{knight1}. There are also some experimental realizations
\cite{experiments}. Here the quantum subsystems involved are the
ionic center-of-mass (phonons) and the quantized field (photons).
Considering that the ion is in its electronic groud state, the
Hamiltonian (\ref{eff}) assumes the following form in the
interaction representation
\begin{equation}
H_{\rm{eff}}=-\frac{g^2}{\Delta}f^2(\hat{a}^\dag\hat{a})\hat{b}^\dag
\hat{b},
\end{equation}
which is adequate for producing general phase shifts. Because
$[a^\dag a,H_{\rm{eff}}]=[ b^\dag b,H_{\rm{eff}}]=0$, we may write
the evolution operator $U(t)=\exp(-iH_{\rm{eff}}t/\hbar)$ in
diagonal form in the basis
$\{|00\rangle,|10\rangle,|01\rangle,|11\rangle\}$, with $|ij\rangle
\equiv |{\rm{ion}}\rangle\otimes|{\rm{field}}\rangle$. Therefore
\begin{eqnarray}
\hat{U}(t)&=&\exp\left\{it\frac{g^2e^{-\eta^2}}{\Delta}[:J_0(\mathfrak{n}):]^2\hat{b}^\dag\hat{b}\right\}\nonumber\\
&=&e^{i\phi_{00}}|00\rangle\langle
00|+e^{i\phi_{10}}|10\rangle\langle
10|+e^{i\phi_{01}}|01\rangle\langle
01|+e^{i\phi_{11}}|11\rangle\langle 11|,
\end{eqnarray}
where
\begin{eqnarray}
\phi_{00}&=&0\nonumber\\
\phi_{10}&=&0\nonumber\\
\phi_{01}&=&\Omega t \nonumber\\
\phi_{11}&=&\Omega t [L_1(\eta^2)]^2,
\end{eqnarray}
being $L_1(\eta^2)$ the first order Laguerre polynomial and $\Omega=-g^2e^{-\eta^2}/\Delta$.
In order to have a $\pi$ phase gate we need to have $\Omega t=2\pi$ and $\phi_{01}-\phi_{11}=-\pi$.
This would demand $[L_1(\eta^2)]^2=1/2$, or $\eta\approx 0.54$,
which is close to realistic values for the Lamb-Dicke parameter
found in the literature \cite{blatt}.

It is now necessary to give a brief discussion on the experimental
implementation of the gate. According to recent experimental data
\cite{blatt3}, the transition from the state $P_{3/2}$ to $D_{5/2}$
in $^{40}$Ca$^{+}$ ions is coupled to the vacuum field with strength
$g=2\pi\times 1.51$ MHz. The losses are mainly due to the cavity at
rate $\kappa=2\pi\times 41.7$ KHz and atomic spontaneous emission at
rate $\gamma=2\pi\times 1.58$ MHz. Choosing $\Delta=10g$, the gate
operation time is $t_{op}\approx 9\,\mu$s. We must compare it to the
decay rates. Firstly, quite remarkable is the fact that the gate
works all time with the ion in the ground state. For this reason the
proposal is quite insensitive to spontaneous emission and this is
much desirable. Secondly, when comparing it to
$\kappa^{-1}=3.8\,\mu$s, one can see that they are at the same order
and so that it is still necessary to improve the quality of the
actual cavities. However, rather than improving the cavity finesse,
one could think of using atomic levels which would provide a
stronger coupling $g$. The important point to be noticed here is
that the robustness against spontaneous emission in our scheme is
the key to improve the coupling constant $g$. We remember that this
coupling is dependent on the atomic decay rate as
$g\backsim\sqrt{\gamma}$ and then one could choose fast decaying
excited levels leading to stronger couplings and consequently faster
gates. It is worthwhile to notice that a phase gate in the
Lamb-Dicke regime [see Eq.(\ref{gs})] would be much faster than the one
we have just discussed.
In that case, $t_{op}\approx 0.62$ms which means
that it is possible to have about 160 cycles per $\kappa^-1$, making
our proposals even more interesting for practical implementation.
Having taken those typical experimental values into account, and
considering the robustness against spontaneous emission, we find
that our proposal opens up new possibilities for feasible quantum
computing and also brings the interesting possibility of
implementing cross-Kerr interactions between bosons in a well
developed experimental system.
\begin{acknowledgments}
The authors would like to thank P.P. Munhoz for a critical reading
of the manuscript. This work is partially supported by CNPq
(Conselho Nacional para o Desenvolvimento Cient\'\i fico e
Tecnol\'ogico), and FAPESP (Funda\c c\~ao de Amparo \`a Pesquisa do
Estado de S\~ao Paulo) grant number 02/02715-2, Brazil.
\end{acknowledgments}


\begin{thebibliography}{xxxxx}

\bibitem{imoto1} N. Imoto, H.A. Haus, and Y. Yamamoto, \pra {\bf 32}, 2287 (1985).

\bibitem{poizat1} P. Grangier, J.A. Levenson, and J.P. Poizat, Nature (London) {\bf 396}, 537 (1998).

\bibitem{dariano1} G.M. D'Ariano, L. Maccone, M.G.A. Paris, and M.F. Sacchi, \pra {\bf 61}, 053817 (2000).

\bibitem{zoller1} L.M. Duan, G. Giedke, J.I Cirac, and P. Zoller, \prl {\bf 84}, 4002 (2000).

\bibitem{tombesi1} D. Vitali, M. Fortunato, and P. Tombesi, \prl {\bf 85}, 445 (2000).

\bibitem{tombesi2} S. Pirandola, S. Mancini, D. Vitali and P. Tombesi, Europhys. Lett. {\bf 68}, 323 (2004).

\bibitem{munro1} K. Nemoto, and W.J. Munro, \prl {\bf 93}, 250502 (2004).

\bibitem{barret} S. D. Barrett, Pieter Kok, Kae Nemoto, R. G. Beausoleil, W. J. Munro,
and T. P. Spiller, \pra {\bf 71}, 060302 (2005)

\bibitem{hau} L. V. Hau, S. E. Harris, Z. Dutton, and C. H. Behroozi, Nature (London) {\bf 397}, 594 (1999).

\bibitem{imamoglu} A. Imamoglu, H. Schmidt, G. Woods, and M. Deutsch, \prl {\bf 79}, 1467 (1999).

\bibitem{tombesi3} C. Ottaviani, D. Vitali, M. Artoni, F. Cataliotti, and P. Tombesi, \prl {\bf 90}, 197902 (2003).

\bibitem{brauns1} J. Zhang, K. Peng, and S.L. Braunstein, \pra {\bf 68}, 035802 (2003).

\bibitem{first} H. Zeng and F. Lin, Phys. Rev. A {\bf 50}, R3589 (1994); V. Bu\v{z}ek, G. Drobn\'y,
M.S. Kim, G. Adam, and P.L. Knight, Phys. Rev. A {\bf 56}, 2352 (1998).

\bibitem{ours1} F.L. Semi\~ao, A. Vidiella-Barranco and J.A. Roversi, Phys. Rev. A
{\bf 64}, 024305 (2001).

\bibitem{ours2} F.L. Semi\~ao, A. Vidiella-Barranco and J.A. Roversi,
Phys. Lett. A, {\bf 299}, 423 (2002).

\bibitem{blatt1} A.B. Mundt, A. Kreuter, C. Becher, D. Leibfried, J. Eschner,
F. Schmidt-Kaler, and R. Blatt, Phys. Rev. Lett. {\bf 89}, 103001 (2002).

\bibitem{blatt2} A.B. Mundt, A. Kreuter, C. Russo, C. Becher, D. Leibfried, J. Eschner,
F. Schmidt-Kaler, and R. Blatt, App. Phys. B {\bf 76}, 117 (2003).

\bibitem{bana} K. Banaszek and K. W\'odkiewicz, \pra {\bf 55}, 3117
(1997).

\bibitem{blatt3} C. Maurer, C. Becher, C. Russo, J. Eschner and R. Blatt, New J. Phys. 6, 94 (2004).

\bibitem{solano1} E. Solano, M. Fran\c ca Santos, and P. Milman, \pra {\bf 64}, 024304 (2001).

\bibitem{scully1} M.S. Zubairy, M. Kim, and M.O. Scully, \pra {\bf 68}, 033820 (2003).

\bibitem{knight1} J. K. Pachos and P.L. Knight, \prl {\bf 91}, 107902 (2003).

\bibitem{experiments} Q. A. Turchette, C. J. Hood, W. Lange, H. Mabuchi, and H. J. Kimble,
\prl {\bf 75}, 4710 (1995);
A. Rauschenbeutel, G. Nogues, S. Osnaghi, P. Bertet, M. Brune, J. M. Raimond, and S. Haroche,
\prl {\bf 83}, 5166 (1999); D. Leibfried, B. DeMarco, V. Meyer, D. Lucas, M. Barrett, J. Britton, W. M. Itano,
B. Jelenkovic, C. Langer, T. Rosenband, D. J. Wineland, Nature (London) {\bf 422}, 412 (2003).

\bibitem{blatt} D. Leibfried, R. Blatt, C. Monroe, and D. Wineland,
\rmp {\bf 75}, 281 (2003).



\end{thebibliography}
\end{document}